\documentclass[lettersize,journal]{IEEEtran}
\usepackage{amsmath,amsfonts}
\usepackage{algorithmic}
\usepackage{algorithm}
\usepackage{array}
\usepackage[caption=false,font=normalsize,labelfont=sf,textfont=sf]{subfig}
\usepackage{textcomp}
\usepackage{stfloats}
\usepackage{url}
\usepackage{verbatim}
\usepackage{graphicx}
\usepackage{cite}
\usepackage{enumerate}
\usepackage{makecell}
\usepackage{hyperref}
\usepackage{multirow}
\usepackage{verbatim}
\usepackage{tablefootnote}
\begin{document}

\title{Proactive security defense: cyber threat intelligence modeling for connected autonomous vehicles}
\author{Yinghui Wang, Yilong Ren, Zhiyong Cui, Haiyang Yu


\IEEEcompsocitemizethanks{\IEEEcompsocthanksitem Y. Wang and Z. Cui are with the School of Transportation Science and Engineering, Beihang University, Beijing 102206, China. Y. Ren and H. Yu are with the Beijing Key Laboratory for Cooperative Vehicle Infrastructure Systems and Safety Control, Beihang University, Beijing 102206, China, and also with the Zhongguancun Laboratory, Beijing 100094, China.     \par $Corresponding Author^1$: Haiyang Yu, South Third Street, Shahe Higher Education Park, Changping District, Beijing, 102206. Telephone number: +86-18810953012, 
E-mail: hyyu@buaa.edu.cn \par $Corresponding Author^2$: Yilong Ren, South Third Street, Shahe Higher Education Park, Changping District, Beijing, 102206. 
E-mail: yilongren@buaa.edu.cn} }



\maketitle
\thispagestyle{empty}
\begin{abstract}
Cybersecurity has become a crucial concern in the field of connected autonomous vehicles. Cyber threat intelligence (CTI), as the collection of cyber threat information, offers an ideal way for responding to emerging cyber threats and realizing proactive security defense. However, instant analysis and modeling of vehicle cybersecurity data is a fundamental challenge since its complex and professional context. In this paper, we suggest an automotive CTI modeling framework, Actim, to extract and analyse the interrelated relationships among cyber threat elements. Specifically, we first design a vehicle security-safety conceptual ontology model to depict various threat entity classes and their relations. Then, we manually annotate the first automobile CTI corpus by using real cybersecurity data, which comprises 908 threat intelligence texts, including 8195 entities and 4852 relationships. To effectively extract cyber threat entities and their relations, we propose an automotive CTI mining model based on cross-sentence context. Experiment results show that the proposed BERT-DocHiatt-BiLSTM-LSTM model exceeds the performance of existing methods. Finally, we define entity-relation matching rules and create a CTI knowledge graph that structurally fuses various elements of cyber threats. The Actim framework enables mining the intrinsic connections among threat entities, providing valuable insight on the evolving cyber threat landscape.

\end{abstract}


\section{Introduction}
\IEEEPARstart {C}{onnected} autonomous vehicles (CAVs) are equipped with advanced embedded systems and connected to various external networks for environment perception and decision-making. The CAVs facilitate significant enhancements in efficiency and road safety, playing a crucial part in intelligent transportation systems \cite{feng2022joint,chen2022artificial}. While CAVs introduce a rang of enabling technologies, the increasing connectivity and system complexity give rise to new attack surfaces or security vulnerabilities, rendering vehicles more vulnerable to potential cyber-attacks\cite{limbasiya2022systematic,bendiab2023autonomous}. For example, two hackers remotely hacked into the Jeep Cherokee via an in-vehicle infotainment system \cite{miller2015remote}. They gained the read/write permission of controller area network (CAN), then remotely manipulated the steering wheel, accelerator, windshield wiper, and even diverted the vehicle from its intended driving direction. The remote cyber-attack incident is regarded as a watershed in the automobile industry and has attracted wide social attention. In the past few years, a notable increase in cyberattacks on automotive systems has been observed. It is forecasted that such attacks will become more common in future autonomous vehicles \cite{kukkala2022roadmap}. Successful cyber-attacks could result in the system failure, vehicle operation disruption, privacy disclosure, and economic loss, etc \cite{hataba2022security,mansourian2023deep}. What's more concerned is that the safety features of automotive cyber-physical systems might be compromised. It could cause human damage and even threat national safety \cite{guan2022overview}. Obviously, the cybersecurity issues of CAVs warrant significant attention. 

\par Over the recent years, significant efforts have been conducted on the cybersecurity of vehicles, such as threat analysis and risk assessment (TARA), CAN bus security, authentication protocols security, intrusion detection, and among others \cite{cui2019review,khan2020cyber,pham2021survey}. Unfortunately, the current security protections for CAVs focus mainly on specific scenarios. With the increasing electronic information components and communication interactions, modern vehicles have evolved into highly intricate systems of systems (SoS) \cite{wiecher2020feature}. Developing security measures for each element of complex vehicle systems against cyber-attacks is not realistic. Moreover, due to the diverse and dynamic nature of cyber-attacks, it is a major challenge to timely discover potential cyber threats and devise suitable security measures for CAVs \cite{comert2021change}. Cyber threat intelligence (CTI) mining, as a more proactive and forward-thinking technique, offers critical insights into emerging cyber-attacks or threats \cite{sun2022defining}. By using CTI to discover evidence-based threats, organizations can proactively predict the actions of threat actors and implement effective defense measures \cite{sun2023cyber}. This empowers them to improve the security posture significantly for CAVs. 

\subsection{Motivation}
\par Most of the CTI data are written in natural language. The previous CTI analysis require intensive manual inspection of the cybersecurity information descriptions, which is a  time-consuming task. Therefore, some studies proposed automated CTI extraction methods in the form of indicator of compromise (IOC) from unstructured texts \cite{liao2016acing,zhu2018chainsmith}. Nevertheless, isolated IOC fails to capture the entire panorama of threat events, making it challenging for organizations to grasp a full picture of the incoming threat. In the meantime, deep learning and natural language processing technology are widely adopted in cyber threat intelligence extraction and analysis. Whereas, most of the recent studies merely focused on cybersecurity entity recognition, ignoring the relationships among entities \cite{dionisio2019cyberthreat,gasmi2019information,satyapanich2020casie,li2019self}. Subsequently, Zhao et al. recommended a CTI framework on the basis of the heterogeneous information network to analyze the dependent relationships among IOCs \cite{zhao2020cyber}. Similarly, Gao et al. created a threat intelligence meta schema, then modeled CTI using the heterogeneous information network \cite{gao2020hincti}. This approach aimed to capture diverse types of nodes and intricate relationships among them. Unfortunately, the meta-paths used in these studies are designed manually. As the IOC types are increasing, the number of meta-paths will also grow significantly. Gradually, researchers have focused on the end-to-end joint extraction model to extract CTI entities and mine their semantic relation \cite{li2020knowledge,zuo2022end,guo2021cyberrel}. However, current CTI research faces three key limitations in the CAVs cybersecurity area: 
\par\textit{1) The diversity of cybersecurity entities and relations, along with the increasing demand for cybersecurity and safety fusion analysis.} Complex vehicle systems result in the diversity of cybersecurity entities and relations, thereby making it harder to extract and understand relevant information. Furthermore, cybersecurity events may involve function safety effects, such as vehicle collision, and human injury, etc. The existing studies neglected the interaction between the automobile cyber and the physical world. 
\par\textit{2) The description of automobile CTI information is quite complex, which would result in the low accuracy of CTI mining.} Different from the traditional IOCs (e.g., IP, hash, or email), the description of cybersecurity entity in automobile CTI has no specific rules or patterns. Moreover, there are many cross-sentence relational entities in automotive CTI data. These aspects will affect the precision of CTI information extraction, and might cause the omission of critical threat entities and relationships.  
\par\textit{3) Lack of labeled CTI corpus.}  Obtaining real-world automotive CTI data can be challenging due to the regulation, industry concern, and professionalism, etc. Compared with more mature areas, the field of automotive cybersecurity lacks a large amount of tagged data for model training.
\par We intend to design a whole scheme for automotive CTI modeling to address the above issues. We refine the challenges to achieve the research goals into two aspects: \textbf{\textit{(1) How to thoroughly capture the diverse entities and their relationships, as well as the relationship between cybersecurity and safety? 
and (2) Considering its complexity and cross-sentence context, how can we effectively acquire automobile CTI knowledge?}}
\subsection{Contributions}
\par To combat these challenges, Actim, a cyber threat intelligence modeling framework is proposed to mine and analyze automotive CTI data. Specifically, Actim defines a vehicle security-safety (VSS) ontology model to represent various elements and their relations in the vehicle cybersecurity data. Then, Actim introduces a document-level entity and relation joint extraction model to handle complex entity descriptions and cross-sentence relations, thereby enabling the effectively mining of automobile CTI information. Finally, we construct a knowledge graph of automobile CTI, which merges the interrelationship among diverse elements, to depict a holistic picture of CTI. Moreover, we have created the first document-level CTI corpus in the automobile cybersecurity domain. In short, the main contributions of our paper are depicted below.

\begin{itemize}
    \item \textbf{Automotive CTI semantic modeling.} We propose an automotive CTI conceptual ontology model, VSS. The ontology captures complex dependencies between cyber and physical systems in CAVs. Specifically, it creates a thorough ontology view of the vehicle, component, vulnerability, attack pattern, consequence, and other security entities, as well as relationships among them. The ontology model establishes a unified and standardized way of expressing automotive security and safety knowledge.
    \item \textbf{Document-level CTI corpus.} We collect 908 cyber threat intelligence documents, then adopt the ``BIOES-entity type-relation type-entity role" schema \cite{luo2020neural} to annotate the entity and relation of the automotive CTI data, called Acti corpus. The corpus contains a total of 3678 sentences, 8195 entities, and 4852 entity relations. To our best knowledge, the Acti corpus we created is the first fully manually annotated dataset for CTI modeling in the automotive cybersecurity field.
    \item \textbf{Cyber threat intelligence mining based on cross-sentence context.} We propose the BERT-DocHiatt-BiLSTM-LSTM model, an automotive cyber threat intelligence mining approach based on cross-sentence context. It models threat-related entities and interdependent relationships among them using an end-to-end document-level joint extraction method. This method introduces a hierarchical attention mechanism, i.e., word-level attention and sentence-level attention. It enables the model to focus on more critical information in the case of complex entity description and cross-sentence relations. In total, our model has advanced performance in cyber threat intelligence mining in the automobile field.
    \item \textbf{Automotive CTI knowledge graph.} We present three matching rules for the extraction of entity-relation triplets. Then, we utilize Neo4j to construct an automotive CTI knowledge graph, ActiKG. 
The ActiKG describes various cyber threat information structurally and relationally to effectively fuse automotive CTI knowledge. It depicts a more comprehensive landscape of cyber threats and facilitates thorough analysis of associations in vehicle cybersecurity and safety. 
\end{itemize}
\par The rest of this paper is outlined as follows. Section \uppercase\expandafter{\romannumeral2} offers an examination of relevant prior studies. After that, the foundational concepts and an outline of the threat intelligence modeling architecture are depicted in Section \uppercase\expandafter{\romannumeral3}. Section \uppercase\expandafter{\romannumeral4} delves into the details of the automotive CTI mining approach. Section \uppercase\expandafter{\romannumeral5} details the experiments and performance outcomes of the presented method, as well as the construction of CTI knowledge graph. Finally, Section \uppercase\expandafter{\romannumeral6} summarizes the paper and highlights future research directions.

\section{Related work}

\par In this section, we review previous research related to ontology modeling for automotive cybersecurity, and cyber threat intelligence mining methods.
\subsection{Automotive cybersecurity conceptual ontology}

\par Several ontology models have been specifically developed for the automotive cybersecurity domain. Shaaban et al. presented an ontology model to verify and validate security requirements against vehicle cybersecurity issues \cite{shaaban2019ontology,shaaban2020automated,shaaban2021ontology}. The proposed ontology included the vehicular component, threat, vulnerability, security requirement and property element. They also extended the model to encompass the assets, broadening the scope beyond merely safety-related aspects to a wider perspective. Similarly, Hou et al. proposed a vehicle cybersecurity ontology model, which was specifically created to generate dynamic attack graphs \cite{hou2022ontology}. They abstracted the security-related elements and their relationships, including the asset, vulnerability and attack. In addition, Grimm et al. dedicated to a vehicle and fleet security method, which aimed to tailor security measures with automotive context-aware \cite{grimm2022context}. In their work, an ontology model was proposed to assist achieve the above goal. They defined the vehicle, context, attack occurrence, attack type, vehicle identification number and location term. In fact, cyber-attacks may affect the physical system of vehicles. However, the above studies fail to explore the safety and cybersecurity issues simultaneously.
\par In order to enhance preventive cybersecurity defense capabilities in vehicle domain, it is crucial to acquire collaborative information between safety and security. Venkata et al. introduced an ontology-driven framework for capturing the connections between physical and cyber systems \cite{venkata2018ontology}. The ontology provided the ability to analyze the consequence of cyber-attacks on physical systems. However, this study only demonstrated an ontology case for vehicle-to-infrastructure reference architecture, including the physical component, cyber component, abstract element, vehicle and infrastructure. Subsequently, Cappelli combined existing ontologies in the internet of things, risk, safety and security domain, to develop an integrated security-safety ontology specifically tailored for vehicle systems \cite{cappelli2022semantic}. This study attempted to analyze the causal relationships between security and safety issues in CAVs. However, the proposed ontology fails to capture the interrelationship of diverse components in vehicles, while ignores several essential security element representations. In other words, these ontologies have not reached the detail level required to analyze the real implementation of the vehicle system. Moreover, they are hard to thoroughly describe cyber threats or attacks associated with vehicle components and technologies.

\subsection{Cyber threat intelligence modeling}

\par There are several attempts to extract CTI from open-source cybersecurity texts. Among them, some researchers employed the heterogeneous information network to analyze and model CTI. Specifically, Zhao et al. introduced a novel IOC recognition method, and employed the heterogeneous information network to analyze the dependent relationships among extracted IOCs \cite{zhao2020cyber}. Similarly, Gao et al. utilized the heterogeneous information network to model CTI by defining meta-paths and meta-graphs over node types \cite{gao2020hincti}. Unfortunately, these studies have a main limitations: the meta-paths or meta-graphs are designed manually, and the number of them will grow significantly as the increasing IOC types. 
\par In the traditional IT domain, researchers began to focus on the joint extraction model for mining CTI entities and analyzing their relationships. Li et al. adopted Luo's joint labeling scheme \cite{luo2020neural}and merged the BiLSTM model with a dynamic attention mechanism to propose a BiLSTM-dynamic-att-LSTM model\cite{li2020knowledge}. This novel approach enabled the simultaneous extraction of cyber entities and their relationships. Subsequently, Zuo et al. introduced a BERT-BiLSTM-att-CRF joint extraction framework \cite{zuo2022end}. The BERT model enables it to effectively capture intricate semantic features of CTI data. Guo et al. modeled the cybersecurity concepts extraction problem as a multi-sequence tagging task \cite{guo2021cyberrel}. The Bert-BiGRU-att-CRF scheme was introduced as an effective solution for extracting cyber entities and their relationships. It should be noted that the above methods only focus on sentence-level cyber entities and relations extraction. Nevertheless, there are many cross-sentence relationships and intricate entity descriptions in automotive CTI data. And these current studies are hard to deal with this complex scenario.

\section{Architecture Overview of Actim}

\subsection{Preliminaries}
\par In order to clarify our work more clearly, some preliminary definitions are first given as follows.\\

\textbf{Definition 1 (\textit{Ontology})}:
\par Ontology is a way used to formally express the concepts and their relationships in a particular domain. The ontology comprises 5 mainly primitives: class (concept), relation, function, axiom and instance \cite{tang2014semantic}. Usually, the ontology model is formalized as: $O=(\mathbb{C},\mathbb{R},\mathbb{F},\mathbb{A},\mathbb{I})$, where $\mathbb{C}$ represents the set of classes or concepts; $\mathbb{R}$ denotes the set of relations between classes; $\mathbb{F}$ stands the set of special relations of function classes; $\mathbb{A}$ means the axiom set of constraint classes and relations; and $\mathbb{I}$ represents the set of class instances.


\textbf{Definition 2 (\textit{Knowledge graph})}:
\par The knowledge graph (KG) is a structured semantic knowledge representation that is utilized to depict entities and their interconnected relationships in the actual world \cite{liu2022recent}. In formal terms, the KG is represented as a directed graph $G=(E,R,S)$, where $E$ and $R$ are the set of entities and relations, respectively. $S$ is represented as the set of triples. Each triple is formulated as $S=\{(e_h,r,e_t)| e_h,e_t\in E,r\in R\}$, where $e_h$, $r$ and $e_t$ denote the head entity, relation, and tail entity, respectively. Entities are the essential elements in the knowledge graph, and relationships link different entities to form a graph structure.


\subsection{System Architecture}
\par Actim, as an automotive cyber threat intelligence modeling framework, is capable of mining entities and their relations from threat-related descriptions. It integrates diverse and heterogeneous cybersecurity data into semantically interpretable knowledge. As depicted in Fig. \ref{fig11}, Actim is composed of 3 primary parts:
\begin{figure*}
\centering
\includegraphics[width=5in]{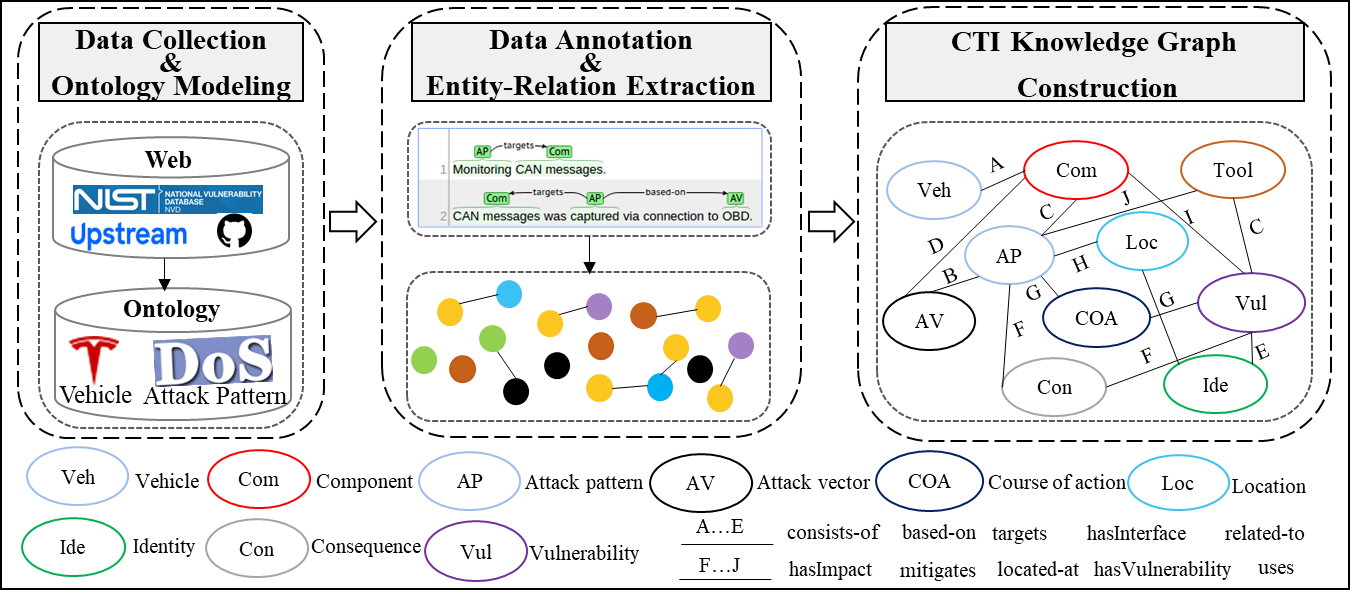}
\caption{The system architecture of Actim. Actim includes three main components: (a) collecting automotive cybersecurity data and modeling CTI ontology; (b) annotating CTI corpus and extracting threat entities and their relationships; and (c) constructing an automotive CTI knowledge graph by leveraging Neo4j.} 
\label{fig11}
\end{figure*}

\begin{itemize}
    \item \textbf{\textit{Data collection and ontology modeling.}} We first collect automotive security-related data from vulnerability databases, public threat intelligence platforms, and research literature, etc. After that, we design a security-safety conceptual ontology model to describe the interdependencies between cyber and physical systems in vehicles. This ontology helps to understand how cybersecurity issues affect physical systems safety (\textbf{see Section \uppercase\expandafter{\romannumeral4.A} for details}).
    \item \textbf{\textit{Cyber threat entity-relation joint extraction.}} The BERT-DocHiatt-BiLSTM-LSTM model addresses the challenge of automotive CTI mining through utilizing hierarchical attention mechanisms and cross-sentence context. The scheme can effectively capture threat entities and their interdependent relationships. In addition, we annotate the collected automotive CTI data using the brat tool to construct the document-level corpus (\textbf{see Section \uppercase\expandafter{\romannumeral4.B} for details, data collection and annotation see Section \uppercase\expandafter{\romannumeral5.A}}).
    \item \textbf{\textit{Knowledge graph construction.}} We present three rules for matching entity-relationship triplets. Subsequently, we leverage Neo4j to construct an automotive CTI knowledge graph, merging various threat-related elements obtained from our previous efforts. A thorough landscape of cyber threats possesses the ability to reveal the evolution trends of vehicle cybersecurity, and further analyze the correlation between vehicle cybersecurity and safety. (\textbf{see Section \uppercase\expandafter{\romannumeral5.C} for details}).
\end{itemize}

\section{Methodology}

\subsection{CTI ontology model construction}
\par A vehicle security-safety conceptual ontology is created through the abstraction of the related elements and their relationships from CTI data. The VSS ontology model encompasses physical and security elements such as the vehicle, component, vulnerability, attack and consequence, etc. It provides a more overall description of vehicle security and safety information.

\subsubsection{\textbf{\textit{Entity definition of the VSS ontology}}}
\par
\begin{figure}
\centering
\includegraphics[width=3.5in]{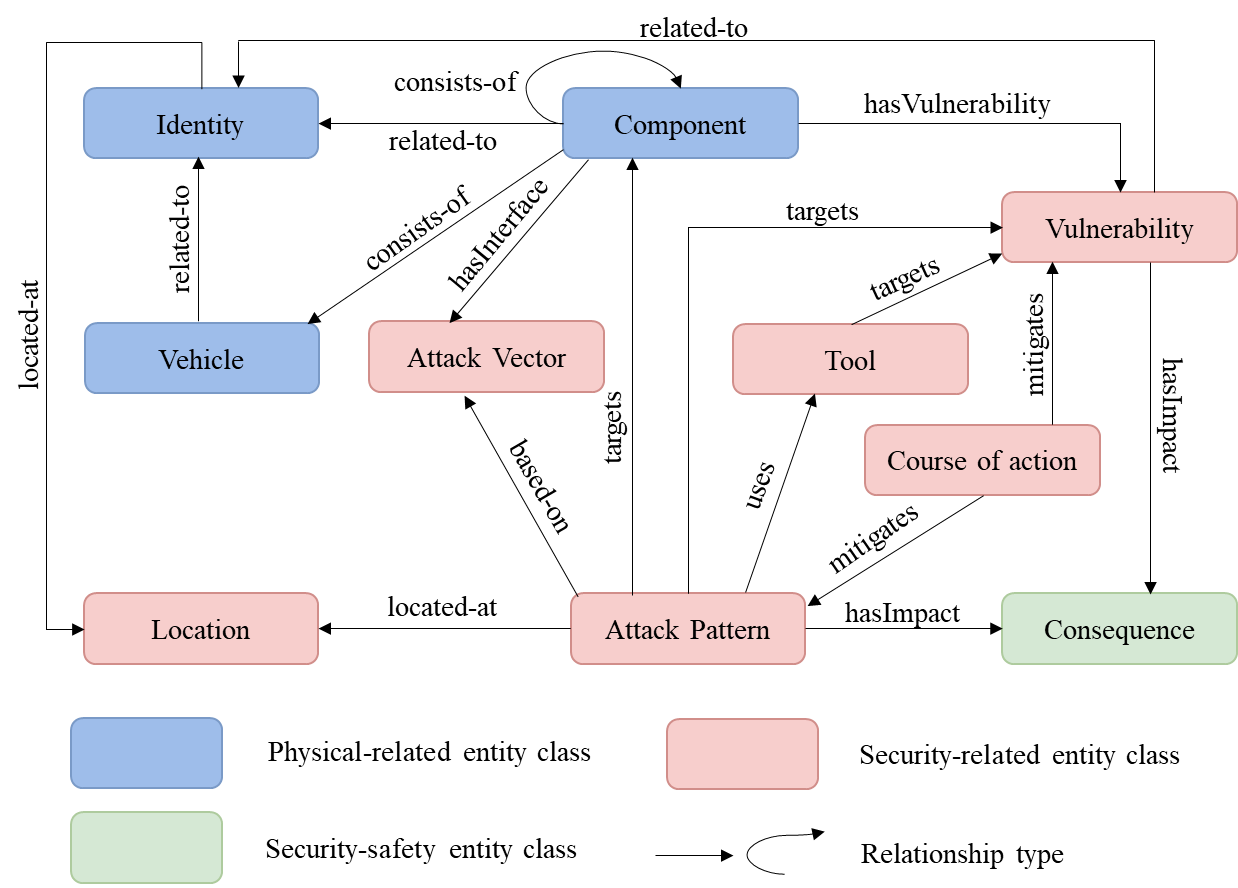}
\caption{Vehicle security-safety conceptual ontology.} 
\label{fig1}
\end{figure}

 The VSS does not intend to be a complete ontology for vehicle cybersecurity. Our goal is to create a core conceptual ontology, as indicated in Fig. \ref{fig1}. The CTI ontology mainly comprises 10 types of entity classes: vehicle, identity, component (asset), attack vector, attack pattern, tool, vulnerability, consequence, location, course of action. The entity classes are defined as follows.

\begin{itemize} 
    \item \par \textbf{Physical-related entity classes}:
\end{itemize}
\par \textit{Vehicle class.} This class refers to the vehicle model or type, e.g., the Tesla Model S, Mercedes-Benz C-Class.
\par \textit{Component class.} This class refers to any element or module of the vehicle system and network, such as software, hardware, services, firmware and data, etc.  
\par \textit{Identity class.} This class represents companies, organizations or groups, e.g., Acura, BMW and Chrysler, etc.

\begin{itemize}
\item \par \textbf{Security-related entity classes}:
\end{itemize}
\par \textit{Attack vector class.} This class is also called attack surface. The main subclasses include cellular networks (3/4/5G), Wi-Fi, Bluetooth, CAN bus, cloud platforms, GPS and keyless entry systems, etc.
\par \textit{Attack pattern class.} This class represents a sequence of steps or activities undertaken by attackers to compromise assets. It is divided into subclasses such as replay, eavesdropping, tampering and buffer overflow, etc.
\par \textit{Vulnerability class.} This class illustrates the logical or technical flaws that might be exploited for launching attacks. For example, buffer overflows, injection flaws, cross-site scripting, misconfiguration, and so on.
\par \textit{Tool class.} This class describes the software or hardware equipment that could be used to execute cyber attacks, such as vehicle diagnostic tools, CANoe, Wireshark, and so forth.
\par \textit{Location class.} The class describes the region where the attack occurred or the target organization/company belongs to.
\par \textit{Course of action class.} The class shows the actions that are taken to protect against cyber-attacks, including intrusion detection, access control, encryption and firewall, etc.
\begin{itemize} 
    \item \par \textbf{Security-safety entity class}:
\end{itemize}
\par
\begin{figure}[htbp]
\centering
\includegraphics[width=2.2in]{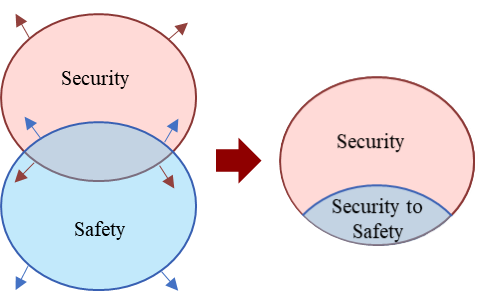}
\caption{Intersection of the security and safety domains.} 
\label{fig22}
\end{figure}

 \textit{Consequence class.} The intersection of the security and safety domains is as shown in Fig. \ref{fig22}. Consequence class is the key elements of the VSS ontology. It describes the potential impact of cybersecurity in terms of financial, privacy, operation, and safety aspects. For example, vehicle theft, control vehicle systems, data breach, service/business disruption, fraud, vehicle collision, and so on.

\subsubsection{\textbf{\textit{Relationship model between entity classes}}}
\par The semantic relations between entity classes mentioned above are reflected through object properties. The following are the object properties defined in the VSS ontology.\\
\textbf{hasVulnerability property:} This property represents that a specific component has a vulnerability instance.
\begin{center}
$\textit{hasVulnerability (Component, Vulnerability)}$
\end{center}
\textbf{hasInterface property:} This property denotes that a component instance has an attack vector instance.
\begin{center}
$\textit{hasInterface (Component, Attack vector)}$
\end{center}
\textbf{hasImpact property:} This property suggests that the attack pattern or vulnerability can cause some impact directly or indirectly. The relationship can be depicted using two triples.
\begin{center}
$\textit{hasImpact (Attack pattern, Consequence)}$
\end{center}
\begin{center}
$\textit{hasImpact (Vulnerability, Consequence)}$
\end{center}
\textbf{targets property:} This property indicates that the attack pattern/tool targets the component, or vulnerability. The relationship can be described using three triples.
\begin{center}
$\textit{targets (Attack pattern, Component)}$
\end{center}
\begin{center}
$\textit{targets (Attack pattern, Vulnerability)}$
\end{center}
\begin{center}
$\textit{targets (Tool, Vulnerability)}$
\end{center}
\textbf{uses property:} This property represents that the tool is employed to carry out the behavior identified in the attack pattern.
\begin{center}
$\textit{uses (Attack pattern, Tool)}$
\end{center}
\textbf{mitigates property:} This property indicates that the course of action can mitigate the related attack pattern or vulnerability. The relationship can be described in two triples.
\begin{center}
$\textit{mitigates (Course of action, Attack pattern)}$
\end{center}
\begin{center}
$\textit{mitigates (Course of action, Vulnerability)}$
\end{center}
\textbf{related-to property:} This property represents that two entity classes have a related relationship. The relationship can be represented in three triples.
\begin{center}
$\textit{related-to (Component, Identity)}$
\end{center}
\begin{center}
$\textit{related-to (Vehicle, Identity)}$
\end{center}
\begin{center}
$\textit{related-to (Vulnerability, Identity)}$
\end{center}
\textbf{located-at property:} This property represents that the identity is located at the related location, or an attack occurred in a region. The relationship can be depicted in two triples.
\begin{center}
$\textit{located-at (Identity, Location)}$
\end{center}
\begin{center}
$\textit{located-at (Attack pattern, Location)}$
\end{center}
\textbf{based-on property:} This property describes that the attack pattern instance was carried out through the attack vector.
\begin{center}
$\textit{based-on (Attack pattern, Attack vector)}$
\end{center}
\textbf{consists-of property:} This property represents that two entity classes have a containment relationship. The relationship can be described in two triples.
\begin{center}
$\textit{consists-of (Component, Vehicle)}$
\end{center}
\begin{center}
$\textit{consists-of (Component, Component)}$
\end{center}

\subsection{Cyber entity-relation joint extraction model}
\par
\begin{figure}
\includegraphics[width=2.8in]{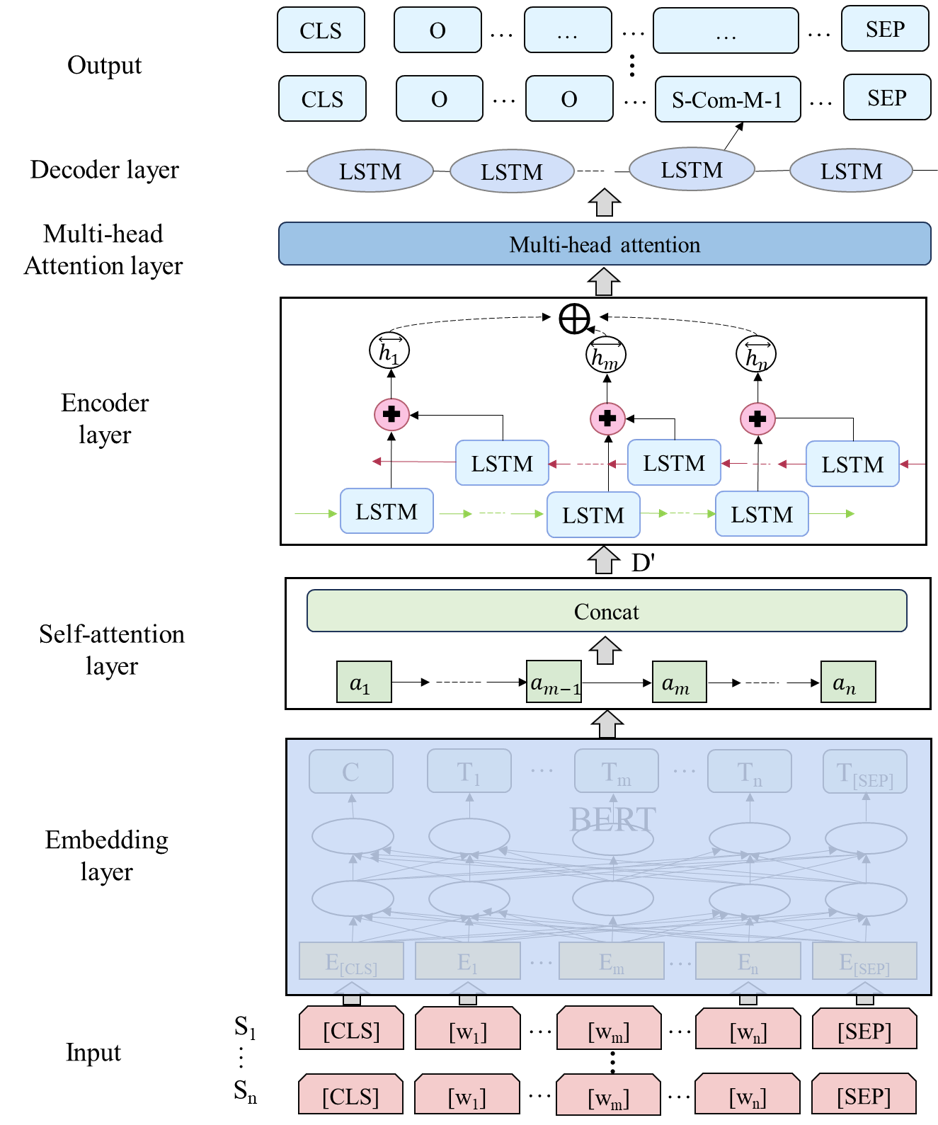}
\caption{BERT-DocHiatt-BiLSTM-LSTM model.} 
\label{fig3}
\end{figure}

 This paper proposes the BERT-DocHiatt-BiLSTM-LSTM model, to achieve cyber threat intelligence mining based on cross-sentence context. As illustrated in Fig. \ref{fig3}, the model is composed of the embedding layer, hierarchical attention mechanism, encoder layer and decoder layer. Given the input sentence {\small$S=\{w_1, w_2, w_3,..., w_n\}$} (with n words) from the document {\small$D=\{S_1, S_2, S_3,..., S_n\}$} (with n sentences), the goal is to extract threat entities and relations by leveraging the context information in document {\small$D$}. Next, we provide a detailed description of each component in this section.

\begin{enumerate}[(1)]
\item \textbf{Embedding layer}
\end{enumerate}
\par For capturing the semantic information of words, each word is converted into a low-dimensional vector. The pre-trained BERT model \cite{devlin2018bert} can fuse contextual information, thus better expressing the semantic features of the words. In the embedding layer, we utilize the BERT model to generate the low-dimensional, dense word embedding vector. Given the limitation of input length for the BERT model, we let the sentence sequence as input. For the input sentence {\small$S=\{w_1, w_2, w_3,..., w_n\}$}, the BERT model processes it to yield the embedding vector {\small$E=\{e_1, e_2, e_3,..., e_n\}$}, where $w_n$ represents the n-th word in the sentence, $e_n$ denotes the word vector of the n-th word. 
\begin{enumerate}[(2)]
\item \textbf{Hierarchical attention mechanism}
\end{enumerate}
\par \textit{Self-attention layer (word-level attention layer)}: Self-attention can learn the word dependencies within sentences, and compute the attention score between individual words. The self-attention mechanism is described as:
\begin{equation}
\small
Attention(Q,K,V)=softmax(\frac{QK^T}{\sqrt{d_k}})V
\end{equation}
where {\small$Q$}, {\small$K$} and {\small$V$} denote the vector matrix of input words, which are derived from sequences that undergo three different mapping transformations. {\small$QK^T$} is the score obtained from the dot product of the {\small$Q$} and {\small$K$}. It indicates the degree of focus given to different segments of a sentence during the encoding of a particular word. $\sqrt{d_k}$ represents the dimension of the input vector. Dividing the {\small$QK^T$} by the $\sqrt{d_k}$ helps to avoid the rise of the score as the increasing input dimensions. 
\par \textit{Multi-head attention layer (sentence-level attention layer)}: In order to capture rich features, a multi-head mechanism is introduced. Essentially, multi-head self-attention conducts multiple self-attention operations simultaneously. It can capture features of different scenes and levels from multi-subspaces \cite{qiao2022event}. The multi-head attention mechanism is expressed as:
\begin{equation}
\small
MultiHead(Q,K,V)= Concat({head}_1, {head}_2,..., {head}_n)
\end{equation}
\begin{equation}
\small
{head}_i= Attention(QW_{i}^{Q}, KW_{i}^{K},..., VW_{i}^{V})
\end{equation}
where n is the number of headers, {\small$W_{i}^{Q}$, $W_{i}^{K}$} and {\small$W_{i}^{V}$} are parameter matrices. After obtaining word embedding representations with global semantic information using the BERT model, this paper applies the attention mechanism to process these embedding representations for further capturing key semantic features. Then, concatenate the BERT feature vectors of the sentence sequences processed by the attention mechanism as the vector representation of the document, {\small$D'=\{E'_1, E'_2, E'_3,..., E'_n\}$}. Where {\small$E'_n$} is the embedding vector of the n-th sentence sequence in the document. To better capture contextual features between sentences and enhance the semantic representation of cross-sentence relational entities in CTI corpora, the multi-head attention mechanism is also employed after the encoder layer.

\begin{enumerate}[(3)]
\item \textbf{Encoder layer}
\end{enumerate}
\par The BiLSTM, or bidirectional long short-term memory, utilizes both forward and backward LSTM \cite{greff2016lstm} to effectively capture semantic information in both directions of each sequence. It could address the issue of gradient disappearance and enhances the generation of comprehensive semantic features. The BiLSTM can be expressed as:
\begin{equation}
\small
\overrightarrow{h_{t}}=LSTM(x_t,\overrightarrow{h_{t-1}})
\end{equation}
\begin{equation}
\small
\overleftarrow{h_{t}}=LSTM(x_t,\overleftarrow{h_{t+1}})
\end{equation}

\begin{equation}
\small
h_t=[\overrightarrow{h_{t}};\overleftarrow{h_{t}}]
\end{equation}
where $x_t$ denotes the input at time step $t$, $\overrightarrow{h_{t}}$ and $\overleftarrow{h_{t}}$ represent forward LSTM output, backward LSTM output at time step $t$, respectively (see Ref.\cite{greff2016lstm} for details of LSTM). Assuming the input vector at the current time is {\small$D'_t$}, and the hidden state obtained from the previous time step is denoted as {\small$h_{t-1}$}, the features learned considering the contextual information at the current time can be represented as:

\begin{equation}
\small
h_t=BiLSTM(D'_t,h_{t-1})
\end{equation}
\par Next, we let the encoded output of BiLSTM as the input for the multi-head attention mechanism. This allows the model to further capture contextual features between sentences.


\begin{enumerate}[(4)]
\item \textbf{Decoder layer}
\end{enumerate}

\par Many entity extraction models have adopted conditional random field (CRF) as the label decoder. However, when the label space is extensive, the CRF-based decoding process can become slow. In contrast, using LSTM as the decoder can significantly accelerate model training and achieve performance comparable to CRF. Thus, faced with the rich label spaces in sequence labeling, this model employs the LSTM network as the decoding layer to obtain the label sequence. In addition, we adopt a biased objective function \cite{zheng2017joint} to enhance the effect of entity tags and reduce the impact of ``O" tag. This objective function is designed to bias towards relational labels, thereby enhancing the association between related entities.


\section{Experimental Evaluation}
\subsection{Dataset and experimental setting}
\par
\begin{figure}[htbp]
    \centering
    \subfloat[{\fontsize{8}{17}\selectfont\fontfamily{ptm}\selectfont Brat annotation}]{%
        \includegraphics[width=0.43\textwidth]{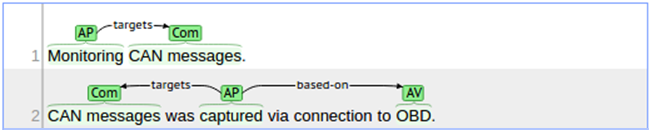}
        \label{fig:sub1}
    }
    \hfill 
    \subfloat[{\fontsize{8}{17}\selectfont\fontfamily{ptm}\selectfont Label CTI corpus}]{%
        \includegraphics[width=0.43\textwidth]{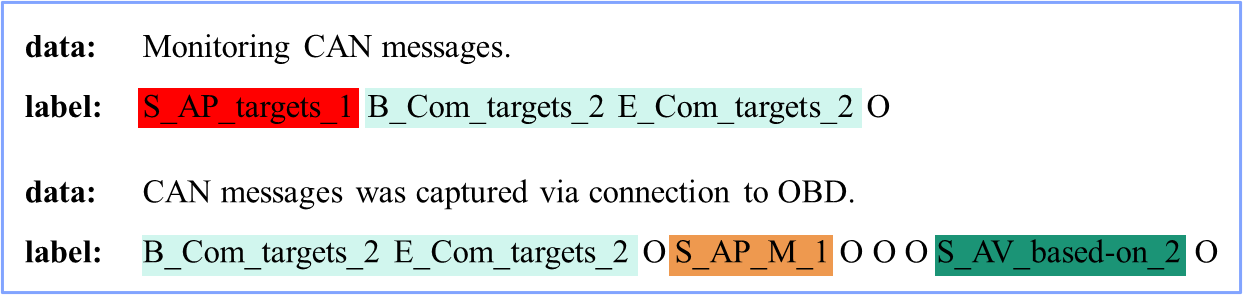}
        \label{fig:sub2}
    }
    \caption{Brat manual annotation.}
    \label{fig:fig5}
\end{figure}

 We collect automotive cybersecurity data from a set of sources, such as vulnerability databases (i.e., national vulnerability database, NVD), threat intelligence platforms (e.g., auto-threat intelligence cyber incident repository of Upstream Security) and GitHub (i.e., automotive attack database, AAD). This paper primarily focuses on analyzing threats or vulnerabilities that directly impact vehicles. We exclude any ransomware attacks, information leakages or other security incidents that occurred in manufacturers backend platforms, car rental platforms, and more. So far, we have collected a total of 908 description texts of automotive cyber threats. To train and evaluate the proposed BERT-DocHiatt-BiLSTM-LSTM model, the collected texts are annotated by using the ``BIOES-entity type-relation type-entity role" schema (see literature \cite{luo2020neural} for details) and VSS ontology model. The annotation data is shown in Fig. \ref{fig:fig5}. 

\par We leverage the open-source Brat tool to effectively label entity and relation types in automotive CTI data. The entity and relation annotation is completed manually, as depicted in Fig. \ref{fig:sub1}. Once the manual annotation process is finished, the Brat tool would automatically generate ``.ann" and ``.conll" files. The ``.conll" file is only available in the standard BIO format. Thus, we develop a conversion script to translate all ``.ann" files into the above annotation format. Take the sequence of ``Monitoring CAN message.", the corresponding label would be represented as ``S-AP-targets-1 B-Com-targets-2 E-Com-targets-2 O". We annotated 908 document-level corpus, which consists of 3678 sentences, 8195 labeled entities and 4852 relationships. All source code for format conversion and preprocessing of the automotive CTI corpus, as well as the complete dataset have been uploaded to GitHub at \href{https://github.com/AutoCS-wyh/Automotive-cyber-threat-intelligence-corpus}{https://github.com/AutoCS-wyh/Automotive-cyber-threat-intelligence-corpus}.
\par In this paper, the training and test samples are randomly divided in a ratio of 8: 2. Based on the CTI corpus, the performance evaluation of the proposed BERT-DocHiatt-BiLSTM-LSTM model is carried out. We run the experiments on Intel Core i5-13400F CPU@2.50 GHz with 64 GB RAM, and the powerful NVIDIA GeForce RTX 3090 GPU. The PaddlePaddle-GPU framework is used to execute the algorithm programs. The hyperparameters settings are detailed in Table \ref{tab:table2}. The proposed model uses the bert-base-cased version of the BERT pre-trained model, which includes 12 transformer layers with a hidden size of 768. The hidden neuron size of the BiLSTM and LSTM networks is 800. To alleviate the overfitting issue during training, this paper introduces the dropout mechanism to enhance the model's generalization ability, with a dropout rate set to 0.1. In the hierarchical attention mechanism, the number of heads for multi-head attention is 8. Moreover, the bias weight in the model's loss function is set to 15. Meanwhile, the adaptive moment estimation (Adam) algorithm is adopted for optimization during model training, with a learning rate set to 2e-5 and the number of training epochs set to 180.

\begin{table}[htbp]
\caption{Hyperparameters information\label{tab:table2}}
\centering
\begin{tabular}{cc}
\Xhline{1pt}
{\textbf{Parameter name}} &{\textbf{Parameter Value}}\\
\Xhline{1pt}
transformer layers & 12 \\
hidden size  & 768\\
activation function& ReLU \\
max position embedding & 128 \\
dimension of BiLSTM encoding & 800  \\
dimension of LSTM decoding &800\\
dropout rate & 0.1 \\
learning rate & 2e-5  \\
nhead & 8 \\
epoch& 180 \\
bias& 15 \\
\Xhline{1pt}
\end{tabular}
\end{table}
\subsection{Results and analysis}
\par To demonstrate the effectiveness of the presented BERT-DocHiatt-BiLSTM-LSTM model, we compare it with the state-of-the-art sentence-level CTI joint extraction models in the traditional IT domain, including BiLSTM-dynamic-att-LSTM \cite{li2020knowledge}, and Bert-BiLSTM-att-CRF \cite{zuo2022end}. In addition, we also compared our model with the BERT-DocHiatt-BiLSTM-CRF method to further evaluate the performance of the CRF and LSTM decoders. The common evaluation metrics of the knowledge extraction task, i.e., precision (P), recall (R) and F1 score, are used to evaluate the performance of the cyber threat intelligence mining model. The extraction results are deemed accurate only when the model correctly labels the entity boundary, entity type, and relation type. The results of experiments conducted using different methods on the Acti corpus are illustrated in Table \ref{tab:table3}. Given the limitations of the corpus resources, we conducted document-level entity and relation joint extraction experiments using 5-fold cross-validation, and averaged the evaluation results of the five folds as the final experiment performance of the model.

\begin{table}[htbp]
\caption{Comparison of experimental results\label{tab:table3}}
\centering
\begin{tabular}{>{\centering\arraybackslash}p{0.8cm}>{\centering\arraybackslash}p{4.5cm}>{\centering\arraybackslash}p{0.5cm}>{\centering\arraybackslash}p{0.5cm}>{\centering\arraybackslash}p{0.5cm}}
\Xhline{1pt}
~&\multirow{2}{*}{\textbf{Model}} & \multicolumn{3}{c}{\textbf{Metric}} \\
\cline{3-5}
~&~& P\% & R\% & F1\% \\
\Xhline{1pt}
\multirow{2}{*}{Sentence}&BiLSTM-dynamic-att-LSTM & 46.03 &42.85  & 44.38 \\
~&BERT-BiLSTM-att-CRF & 50.9 & 55.84 & 53.26  \\
\multirow{2}{*}{Document}&BERT-DocHiatt-BiLSTM-CRF& 52.1 & 31.42 & 39.19\\
~&BERT-DocHiatt-BiLSTM-LSTM(bias) & \textbf{54.25} & \textbf{31.79} & \textbf{40.08}\\
\Xhline{1pt}
\end{tabular}
\end{table}

\par In fact, in the field of cybersecurity for connected autonomous vehicles, we prioritize the precision of information extraction models. In other words, we need to ensure that the identified security entities and relationships are as accurate as possible. According to the experimental results, the proposed BERT-DocHiatt-BiLSTM-LSTM model exhibited outstanding performance in extracting automotive cyber threat intelligence information with a precision score of 54.25\%. 
Observably, the model with the LSTM decoder performs better in the document-level entity relation joint extraction task. Furthermore, this model demonstrated faster running speed, with a total runtime of 26h 44m 30s, which is only approximately half that of the BERT-DocHiatt-BiLSTM-CRF model. From the experimental results of BiLSTM-dynamic-att-LSTM and BERT-BiLSTM-att-CRF models, the BERT-based model outperforms the BiLSTM-dynamic-att-LSTM method, which also proves the exceptional capability of BERT in the entity relation extraction tasks. The reason behind this is that BERT word vector better captures grammatical and semantic information across various contexts, improving the model's generalization ability. It is capable of effectively handling the complex semantic characteristics of automotive CTI data. 
\par
\begin{figure}[ht]
\includegraphics[width=3.5in]{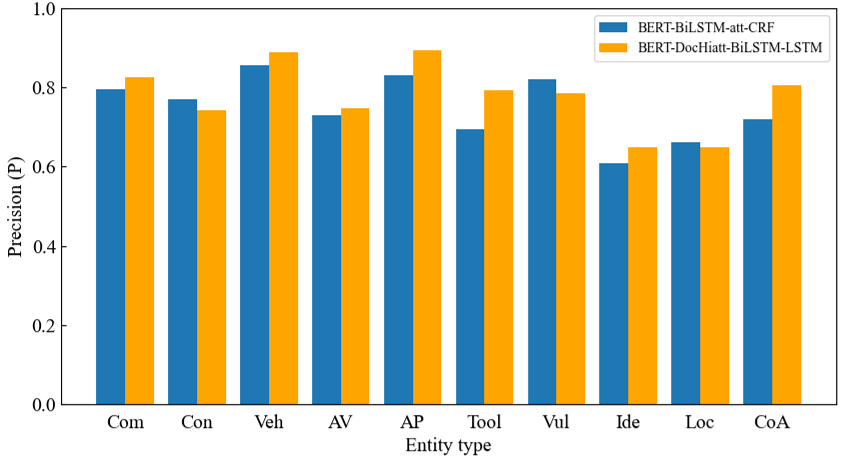}
\caption{Experimental results on entities.} 
\label{fig6}
\end{figure}

\begin{figure}[ht]
\includegraphics[width=3.5in]{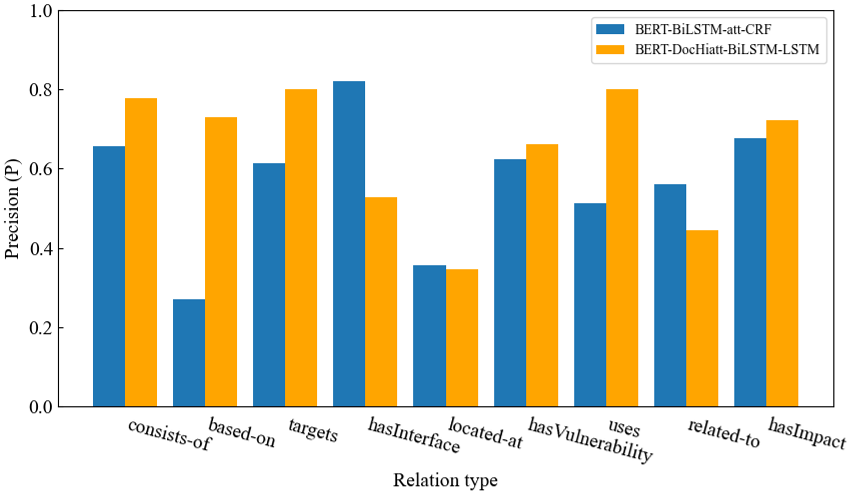}
\caption{Experimental results on relations.} 
\label{fig7}
\end{figure}

 In comparison to BERT-BiLSTM-att-CRF, the precision score of our model improved by 3.35\%. The BERT-DocHiatt-BiLSTM-LSTM model achieves document-level feature extraction by leveraging cross-sentence context information. Furthermore, our approach integrates a hierarchical attention mechanism to learn feature weights of sentence sequences and contextual features between sentences. This enables the model to more effectively characterize the input sequences, thereby further improving the precision of information extraction. To conduct a more detailed evaluation of the model's performance, we analyze the experimental results for each class of entity and relation extraction in both the BERT-BiLSTM-att-CRF and our model, as shown in Fig.\ref{fig6} and Fig.\ref{fig7}.


\par Our CTI mining model has achieved varying degrees of improvement in most entities or relations extraction compared to BERT-BiLSTM-att-CRF, especially the performance of relation extraction has significantly improved. In the automobile CTI corpus, the sample size of several entities and relationships is very small, specifically  ``Course of action", and ``mitigates". The precision score of ``mitigates" relation extraction is almost 0 in both BERT-BiLSTM-att-CRF and BERT-DocHiatt-BiLSTM-LSTM models. Moreover, the CTI corpus contains numerous overlapping relationships, i.e., one entity may be involved in several semantic relationships simultaneously. These overlapping relations might cause ambiguities, thus making it difficult to recognize such entities and relationships. Thus, the CTI mining model proposed in this paper has room for further improvement.

\par The BERT-DocHiatt-BiLSTM-LSTM model introduces a hierarchical attention mechanism to capture the dependencies between words within sentences, as well as the contextual features across sentences. Additionally, a bias function is adopted at the decoder layer to enhance relational connections between entities. To verify the effectiveness of these components in entity-relation joint extraction, we conduct ablation experiments to evaluate their impact on the performance of the model. The experimental results are shown in Table \ref{tab:table4}.

\begin{table}[htbp]
\caption{Effect of attention mechanism and bias function on model performance\label{tab:table4}}
\centering
\begin{tabular}{ccccc}
\Xhline{1pt}
\multirow{2}{*}{\textbf{Model}} & \multicolumn{3}{c}{\textbf{Metric}} \\
\cline{2-4}
~& P\% & R\% & F1\% \\
\Xhline{1pt}
BERT-BiLSTM-att-LSTM & 52.75 & 31.78 & 39.66  \\
BERT-att-BiLSTM-LSTM & 53.56 & 32.15 & 40.18\\
BERT-BiLSTM-Multi-att-LSTM & 53.48 & 32.15 & 40.15\\
BERT-att-BiLSTM-att-LSTM & 53.74 & 31.6 & 39.78\\
BERT-DocHiatt-BiLSTM-LSTM & 52.75 & 31.42& 39.37\\
BERT-DocHiatt-BiLSTM-LSTM (bias) & 54.25 & 31.79& 40.08\\
\Xhline{1pt}
\end{tabular}
\end{table}
\par Clearly, the BERT-DocHiatt-BiLSTM-LSTM model, with the bias function, still has the highest precision score. However, the experimental results indicate that the recall scores of the document-level joint extraction model are relatively low. It is primarily due to the limitations of document-level CTI corpus resources. To address this, we employed the data augmentation method to train the proposed model. The precision, recall, and F1 score achieved by the BERT-DocHiatt-BiLSTM-LSTM (bias) model are 72.14\%, 45.53\%, and 55.83\% respectively. To some extent, this approach can improve the model's performance. Nonetheless, the current data augmentation method is relatively simple. In future research, we will consider optimizing the data augmentation method and annotating more datasets to further expand the automotive CTI corpus.

\subsection{Automotive CTI knowledge graph construction}
\par
\begin{figure}[ht]
\includegraphics[width=3.55in]{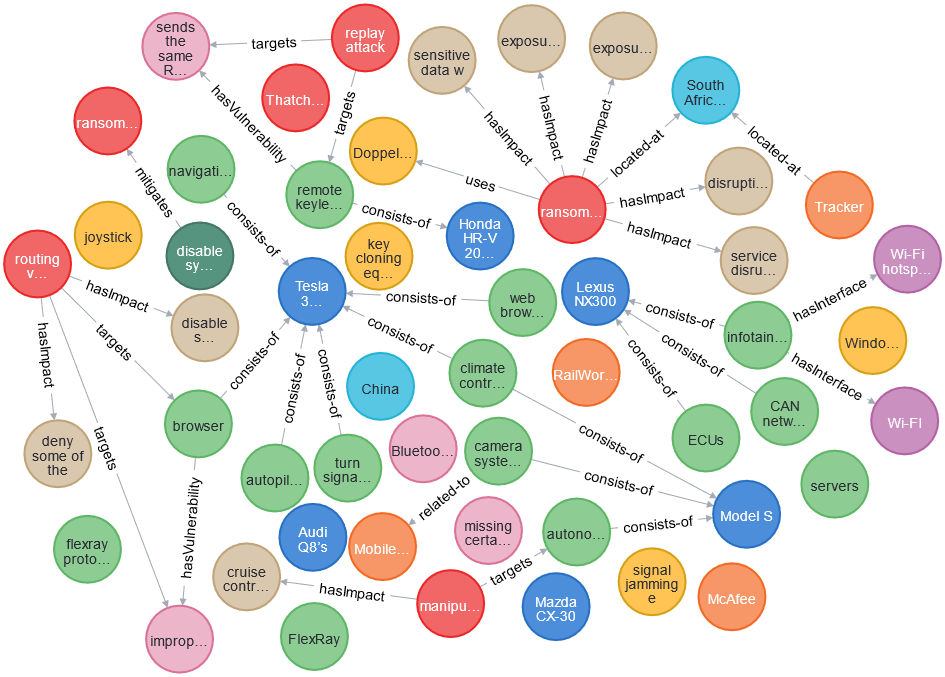}
\caption{A sample of the ActiKG.} 
\label{fig4}
\end{figure}

 We attempt to convert the label sequences predicted through BERT-DocHiatt-BiLSTM-LSTM model into entity-relation knowledge triples. Firstly, the cyber threat entity is obtained based on the entity boundary and type. Then, the knowledge triple is formed based on the relation type and entity role. For this purpose, we define the following three matching rules: (1) The entity is extracted based on labels that define the boundaries and types of tokens. If tokens inside an entity possess different entity type labels, the first token's label is determined as the type label of this entity. Similarly, this principle is also applied to processing relation type and entity role label. (2) Regarding the relation type, entities can be matched only if they possess the same relation type label. Except for entities labeled with the ``M" relation, which can pair with entities of any relation type. Certainly, these matching rules should follow the 18 types of triples outlined in Section \uppercase\expandafter{\romannumeral4.A}. Besides, for the entity role label, an entity can only be paired with a target entity that has a distinct entity role label. (3) Entity role matching follows the bidirectional principle. The entity with entity role label ``1" is looking forward and backward, searching for the entities labeled with ``2" or ``m".  The entity with entity role ``2", one also should search bidirectionally to match entities with entity role ``1" or ``m". When dealing with the entity with entity role ``m", one should bidirectionally traverse any entities for matching. To produce ActiKG, the extracted knowledge triples are processed and outlined in the KG as \textbf{Definition 2}. The extracted triples, denoted as $(e_h, r, e_t)$, consist of a head node $e_h$ and a tail node $e_t$, both of which belong to the entity set {\small$E$}. Two nodes are connected through an edge $r$, representing the relationship between them. A sample of the created ActiKG utilizing Neo4j is demonstrated in Fig. \ref{fig4}. 
\par The ActiKG consists of circle labels and directed edges. The circle labels and their textual descriptions represent cyber threat entities in automotive cyber threat intelligence. It includes information about the vehicle, component, attack pattern, consequence and other entity instances. Different colors represent different entity types predefined in the ontology model, as shown in Table \ref{tab:table1}. The textual descriptions in the middle of the arrows represent the relationship types between different entities, which have also been predefined in Section \uppercase\expandafter{\romannumeral4.A}. The ActiKG links the cyber threat entities extracted from unstructured CIT data, forming a threat intelligence knowledge graph that embodies the characteristics of security in connected autonomous vehicles. Notably, entity disambiguation and completion in KG are separate research problems that are out of the scope of this paper.

\begin{table}[htbp]
\caption{Entity Labels in the ActiKG\label{tab:table1}}
\centering
\begin{tabular}{>{\centering\arraybackslash}p{0.5cm}>{\centering\arraybackslash}p{1.9cm}>{\centering\arraybackslash}p{5.5cm}}
\Xhline{1pt}
{\textbf{Label}} &{\textbf{Type}}&{\textbf{Security threat instance}} \\
\Xhline{1pt} 
\multirow{1}{*}{\includegraphics[width=0.115in]{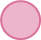}} & Vulnerability & Bluetooth code execution vulnerability \\
\multirow{1}{*}{\includegraphics[width=0.115in]{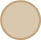}} & Consequence & exposure of personally identifiable information \\
\multirow{1}{*}{\includegraphics[width=0.115in]{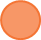}} & Identity & Mobileye Inc \\
\multirow{1}{*}{\includegraphics[width=0.115in]{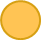}} & Tool & signal jamming equipment \\
\multirow{1}{*}{\includegraphics[width=0.115in]{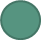}} & Course of action & disable their systems \\
\multirow{1}{*}{\includegraphics[width=0.115in]{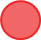}} &Attack pattern & replay attack \\
\multirow{1}{*}{\includegraphics[width=0.115in]{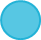}} & Location & South African \\
\multirow{1}{*}{\includegraphics[width=0.115in]{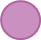}} & Attack vector & Wi-Fi hotspot \\
\multirow{1}{*}{\includegraphics[width=0.115in]{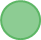}} & Component & remote keyless system \\
\multirow{1}{*}{\includegraphics[width=0.115in]{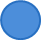}} & Vehicle & Tesla model 3\\
\Xhline{1pt}
\end{tabular}
\end{table}

\section{Conclusions}
\par This paper explores a new direction of proactive security defense on automobiles, which attempts to discover valuable knowledge from numerous cyber threat information. 
The purpose of our work is to capture the relationships between vehicle physical and cyber systems, and then semantically correlate the automotive cybersecurity data to timely identify potential security threats. We propose the Actim, a cyber threat intelligence modeling framework, to mine and analyze automotive CTI data by leveraging the information extraction technology. According to the actual demand for CTI analysis, we design an automotive CTI conceptual ontology, and manually annotate the first document-level CTI corpus based on real-world cybersecurity data. We also propose an end-to-end automotive CTI mining approach, and then construct a CTI knowledge graph to fuse diverse cyber threat elements. The proposed CTI mining framework can effectively extract vehicle security-safety knowledge hidden in numerous cyber threat information. It helps to timely detect potential cyber threats, and formulate appropriate security measures for automobiles. The experiment outcomes show the effectiveness of the CTI mining model presented in this article.


\par In the future, we will attempt to formulate a more detailed operational ontology that represents intricate relationships between diverse elements. Meanwhile, we intend to work on advancing the model's ability to learn the overlapping relational triples. Subsequently, we turn our focus to KG completion or link prediction to improve the strength of the constructed ActiKG. Additionally, the corpus will also be continuously expanded to support further automotive CTI modeling research.

\bibliographystyle{IEEEtran}
\bibliography{ref}

\vfill

\end{document}